\documentclass[12pt]{article}
\usepackage[esperanto,english]{babel}
\usepackage{epsfig}

\usepackage{parallel} 
\newcommand{\ppunu}[1]{}  \newcommand{\ppdu}[1]{#1}

\newcommand{\comentario}[1]{}

\ppdu{
\newlength{\pplw}\setlength{\pplw}{0.49\textwidth}
\newlength{\pprw}\setlength{\pprw}{0.49\textwidth}

\newcommand{\ppp}{\ParallelPar}
\newcommand{\ppn}{\noindent}              
\newcommand{\ppl}[1]{\ParallelLText{\selectlanguage{esperanto}#1}}
\newcommand{\ppr}[1]{\ParallelRText{\selectlanguage{english}#1}\ppp}
\newcommand{\ppln}[1]
{\ParallelLText{\ppn \selectlanguage{esperanto}#1}} 
\newcommand{\pprn}[1]
{\ParallelRText{\ppn \selectlanguage{english}#1}\ppp} 

\newcommand{\ppa}{\addtocounter{section}{-1}}
\newcommand{\ppas}{\addtocounter{subsection}{-1}}
\newcommand{\ppR}[1]{\ParallelRText{#1}}

\newcommand{\ppsection}[3][0ex]{\vspace{2em} 
\ppl{\section{#2} \vspace{#1}} \ppa \nopagebreak
\ppR{\section{#3}} \ppp \nopagebreak}

\newcommand{\ppsubsection}[3][0ex]{\vspace{2em}
\ppl{\subsection{#2}\vspace{#1}} \ppas \nopagebreak
\ppR{\subsection{#3}} \ppp \nopagebreak}

\newcommand{\bea}{\vspace{-1ex}\begin{eqnarray}}
\newcommand{\eea}{\end{eqnarray}}
\newcommand{\dd}{\mathrm d}
}

\ppunu{
\newcommand{\ppl}[1]{\selectlanguage{esperanto}#1}
\newcommand{\ppln}[1]{\noindent \selectlanguage{esperanto}#1}
\newcommand{\ppr}[1]{\selectlanguage{english}}
\newcommand{\pprn}[1]{\selectlanguage{english}}
\newcommand{\ppsection}[3][0ex]{\section{#2}}
\newcommand{\ppsubsection}[3][0ex]{\subsection{#2}}
\newcommand{\bea}{\begin{eqnarray}}
\newcommand{\eea}{\end{eqnarray}}
}

\title{La relativeca tempo -- I \ppdu{\hfill --- \hfill The relativistic time -- I}}
\author{F.M. Paiva \\ 
{\small Departamento de F\'\i sica, Unidade Humait\'a II, Col\'egio Pedro II} \\
{\small Rua Humait\'a 80, 22261-040  Rio de Janeiro-RJ, Brasil; fmpaiva@cbpf.br} 
\vspace{.7ex} \\
A.F.F. Teixeira \\
{\small Centro Brasileiro de Pesquisas F\'\i sicas} \\
{\small 22290-180 Rio de Janeiro-RJ, Brasil; teixeira@cbpf.br}}
\date{14--a de Marto de 2006}

\begin{document}
\selectlanguage{esperanto}
\maketitle

\begin{abstract}\selectlanguage{esperanto}
La relativeca tempo estas malsama ol la Newtona. Ni revidas iujn el tiuj malsamoj en Dopplera efiko, \^gemel-paradokso, rotacio, rigida stango, kaj konstanta propra akcelo.

\ppdu{\selectlanguage{english}
The relativistic time is different from the Newtonian one. We review some of these differences in Doppler effect, twin paradox, rotation, rigid rod, and constant proper acceleration.}
\end{abstract}
\selectlanguage{english}

\ppdu{
\begin{Parallel}[v]{\pplw}{\pprw}
}

\ppdu{\section*{\vspace{-2em}}\vspace{-2ex}}   

\ppsection[0.6ex]{Enkonduko}{Introduction}

\ppln{Astronomiaj observoj anta\u u longe montris, ke la rapido $c$ de lumo en vakuo ne pendas de la rapido de fonto nek de la rapido de observanto; \^ci tiu konstato generis en 1905 la relativecan teorion (specialan) kaj poste la konvencion~\cite{BIPM}: \textsf{la rapido de lumo en vakuo estas senerare} $c:=$ 299.792.458 $m/s$. Tiu konvencio kunigas {\em metron} kun {\em sekundo}, t.e., la ekzisto de la universala konstanto $c$ iele unuigas tempon kun spaco. Tamen, \^ciutage oni preferas distingi la konceptojn de intertempo kaj de interspaco; eble \^car $1 s$, kiu estas intertempo konvena por la homa skalo, samvaloras interspacon tro granda por nia skalo.}
\pprn{Astronomical observations have shown long ago that the speed $c$ of light in vacuum does not depend on the speed of the source, nor on the speed of the observer; this fact gave birth in 1905 to the (special) theory of relativity, and later to the convention~\cite{BIPM}: \textsf{the speed of light in vacuum is exactly $c:=299.792.458m/s$}. This convention ties {\em metre} with {\em second}, i.e., the {\mbox existence} of the universal constant $c$ somehow unites time and space. However, in everyday life one prefers distinguish the concepts of time and space; possibly because $1 s$, which is a time interval convenient for the human scale, {\mbox corresponds} to a distance too large for our scale.}

\ppl{Aliaj eksperimentoj~\cite{Hay} montris, ke la {\em rapido} de luma fonto \^san\^gas la periodon de observata lumo (Dopplera efiko), malgra\u u ke la {\em akcelo} ne \^san\^gas \^gin. Tiuj faktoj forte sugestis la uzon de elektromagneta radiado por difini unuon de intertempo~\cite{BIPM},~\cite[pa\^go~28]{MTW}: \textsf{{\em sekundo} (intervala unuo $\Delta\tau$ de {\em propratempo}) estas la da\u uro de $9.192.631.770$ periodoj de la lumo igita el atomo de cezio-133, je la transiro inter du specifaj niveloj.} El la difinoj de $c$ kaj de {\em sekundo} rezultis, ke {\em metro} estas la distanco ke lumo trakuras en vakuo dum la frakcio $1/299.792.458$ de {\em sekundo}~\cite{BIPM,Salgado}.}
\ppr{Other experiments~\cite{Hay} have shown that the {\em speed} of a light source changes the period of the light observed (Doppler effect), although the {\mbox{\it acceleration}} does not change it. These facts strongly suggested the use of electromagnetic radiation to define a unit of time interval~\cite{BIPM},~\cite[page~28]{MTW}: \textsf{{\em second} (an interval unit $\Delta\tau$ of {\em propertime}) is the duration of $9.192.631.770$ periods of the light emmited from an atom of cesium-133, in the transition between two specific levels.} From the definitions of $c$ and {\em second}, it resulted that {\em metre} is the distance that light traverses in vacuum in the fraction $1/299.792.458$ of a {\em second}~\cite{BIPM,Salgado}.}

\ppl{Horlo\^go fide montranta la sinsekvon de la sekundoj nomi\^gas {\em normohorlo\^go}, a\u u simple {\it horlo\^go} \cite[pa\^go\,107]{Synge2}. La plej fidindaj horlo\^goj estas la ellaborataj atomohorlo\^goj. Tamen la familiara brakhorlo\^go estas anka\u u normohorlo\^go, se ni ne postulas tro precizan mezuron de tempo.}
\ppr{A clock that faithfully shows the succession of the seconds is called a {\em standard clock}, or simply a {\it clock} \cite[page\,107]{Synge2}. The most faithful clocks are the sophisticate atomic clocks. However the popular whristclock is also a standard clock, if we do not demand a too precise time measurement.}

\ppl{Konvenas memori, ke la horlo\^ga takto ne estas absoluta, la\u u la jena senco. Imagu du komence apudajn sinkronajn horlo\^gojn. Poste apartigu kaj submetu ilin al malsamaj kondi\^coj de rapido kaj de gravita potencialo. \^Ciu horlo\^go da\u urigos registri la fluon de {\em sia} propratempo, pere de la akumulado de {\em siaj} sekundoj. En eventuala renkonto de la horlo\^goj, la du akumulitaj registra\^{\j}oj probable estos malsamaj.}
\ppr{It is worth remembering that the pace of a clock is not absolute, in the following sense. Suppose two clocks initially close, and synchronous. Then separate and submit them to different conditions of speed and gravitational potential. Each clock keeps registering the flow of {\em its} propertime, through the accumulation of {\em its} seconds. In an eventual reencounter of the clocks, the two accumulated readings probably will be different.}

\ppl{En la sekvantaj sekcioj ni raportos iujn {\mbox tempajn} fenomenojn \^ce la special-relativeca kunteksto. Ni studos la tempon \^ce eksperimen\-taj rezultoj, Dopplera efiko, \^gemel-paradokso, sin\-kron\-igo kiel Einstein, horlo\^goj en rekta movado, horlo\^goj en cirkla movado, periodo, rigida stango, kaj konstanta propra akcelo.}
\ppr{In the following sections we shall discuss some time phenomena in the special relativistic context. We shall study the time in experimental results, Doppler effect, twin paradox, Einstein synchronization, clocks in rectilinear motion, clocks in circular motion, period, rigid rod, and constant proper acceleration.}

\ppl{Indas citi Rindler~\cite[pa\^go~44]{Rindler}, kiu emfazis ke ``tempa dilato, same kiel spaca maldilato, estas {\em reala}''. Sekve \^ciu, kiun ni diras pri normohorlo\^ga takto, supoze validas por \^cia ajn fizika fenomeno.}
\ppr{Worth citing Rindler~\cite[page~44]{Rindler}, who emphasized that ``time dilation, like length contraction, is {\em real}\,''. As a consequence, all we say about pace of clocks is supposed valid for any physical phenomenon.}
  
\ppsection{Eksperimentaj rezultoj}{Experimental results}

\ppln{Ni komence mencias du eksperimentajn rezultojn kiuj konfirmas specialan relativecon.}
\pprn{We initially mention two experimental results that confirm special relativity.}

\ppsubsection{Muona disi\^go}{Muon disintegration}

\ppln{La malrapideco de movi\^gantaj horlo\^goj estas observata \^ce la defalo de muonoj kiuj estas kreataj en alta tera atmosfero kaj disi\^gas en siaj trajektorioj al detektiloj en malplia altitudo. Ja, la muona meza-vivo estas $\Delta\tau\approx 2,22\mu s$ (propratempo). La\u u la Newtona kinematiko, tiuj muonoj, e\^c se voja\^gus kun rapido $c$, devus trakuri apena\u u \^cirka\u u 700$m$ anta\u u disi\^gi; do nur sensignifa procento de ili devus esti trovata en marnivelo. Tamen, la procento detektita tie estas multe pli granda~\cite[pa\^go~702]{Eisberg}. La relativeca motivo estas, ke muonoj voja\^gantaj kun rapido $v=0,9999c$ vivas \^cirka\u u $70\Delta\tau$, kiel mezurita per horlo\^go ripozanta \^ce grundo. Do tiuj muonoj povas voja\^gi \^cirka\u u $50km$ anta\u u disi\^gi, kio eksplikas \^gian abundan detektadon en anka\u u malgranda altitudo.}
\pprn{The slowness of clocks in motion is observed in the decay of muons that are created in the high terrestrial atmosphere and disintegrate in their path towards detectors in lower altitude. Indeed, the muon lifetime is $\Delta\tau\approx 2,22\mu s$ (propertime). According to the Newtonian kinematics these muons, even running at speed $c$, should travel some 700$m$ only, before disintegrating;  so only a few percent of them should be found at sealevel. Nevertheless, the  quantity detected there is much larger~\cite[page~702]{Eisberg}. The relativistic reason is that muons travelling at speed $v=0,9999c$ live nearly $70\Delta\tau$, as measured by a clock at rest on the ground. So these muons can travel nearly $50km$ before disintegrate, what explains their profuse detection also at low altitude.}

\ppsubsection{Ciklotrono}{Cyclotron}

\ppln{La special-relativeco anka\u u korekte pritraktas la malrapidecon de la evoluo de sistemoj kun rapida cirkla movado en la interno de ciklotrono. Ja, nefirmaj partikloj havante mallongajn meza-vivojn prezentas grandan postvivon \^ce tiuj kondi\^coj~\cite{Bailey}. Tio estas eksplikata denove per la relativeca tempa dilato.}
\pprn{The special relativity deals also correctly the lentitude of the evolution of systems in fast circular motion inside a cyclotron. Really, unstable particles with short lifetimes show a large surviving under these conditions~\cite{Bailey}. That is again explained by the relativistic time dilation.}

\ppsection{Doppleraj efikoj}{Doppler effects} 

\ppln{Imagu ke mi havas \^ce mi horlo\^gon $\cal H$, kiu montras la fluon de mia propratempo $\tau$. Mi estas rigardanta alian horlo\^gon $\cal H'$, kiu montras la fluon de sia propratempo $\tau'$. La horlo\^go $\cal H'$ movi\^gas kun vektora rapido $\bf v$ en inercia referenca sistemo kie $\cal H$ restas. Dum mi {\bf vidas} (nudokule a\u u dulornete a\u u teleskope) ke la montro de $\cal H'$ pligrandi\^gas $\dd\tau'$, la montro de $\cal H$ pligrandi\^gas $\dd\tau_p\neq \dd\tau'$. Ni uzas la subsignon $p$  (post) en $\dd\tau_p$ por memori, ke dum $\dd\tau'$ estas la propra intertempo inter du lumaj eligoj de $\cal H'$\,, kontra\u ue $\dd\tau_p$ estas la propra intertempo inter la alvenoj de tiuj signaloj al $\cal H$. Tiu malsamo de montroj {\it \^ciam} havas relativecan faktoron (``tempan dilaton''), multiplikante la ordinaran nerelativecan Doppleran faktoron. La rilato inter $\dd\tau_p$ kaj $\dd\tau'$ pendas de la modulo de la rapido $\bf v$, kaj de la angulo $\alpha$ inter $\bf v$ kaj la vida rekto $\cal L$ de $\cal H$ al $\cal H'$. Vidu figuron~\ref{FDoppler}.}
\pprn{Suppose I have with me a clock $\cal H$, that shows the flow of my propertime $\tau$. I am looking to another clock $\cal H'$, that shows the flow of its propertime $\tau'$. The clock $\cal H'$ is moving with vector speed $\bf v$ in an inertial reference system where $\cal H$ rests. While I {\bf see} (with naked eyes, or through a binocular, or through a telescope) that the reading of $\cal H'$ increases $\dd\tau'$, the reading of $\cal H$ increases $\dd\tau_p\neq \dd\tau'$. We use the subindex $p$ (post) in $\dd\tau_p$ to remember that while $\dd\tau'$ is the interval of propertime between two light emissions from $\cal H'$\,, $\dd\tau_p$ is the propertime interval between the arriving of these signals to $\cal H$. That difference of readings {\it always} has a relativistic factor (``time dilation''), multiplicating the usual nonrelativistic Doppler factor. The relation between $\dd\tau_p$ and $\dd\tau'$ depends on the modulus of speed $\bf v$, and on the angle $\alpha$ between $\bf v$ and the line of sight $\cal L$ from $\cal H$ to $\cal H'$. See the figure~\ref{FDoppler}.}

\ppl{Nun ni rilatas $\dd\tau'$ al $\dd\tau_p$ \^ce kvin okazoj de Dopplera efiko. Ni vidos ke en tiuj okazoj la relativeca kaj la Newtona anta\u uvidoj koincidas nur en unua ordo de $v/c$.}
\ppr{We now relate $\dd\tau'$ to $\dd\tau_p$ in five cases of Doppler effect. We shall see that in these cases the relativistic and Newtonian predictions coincide only in first order of $v/c$.} 

\begin{figure}                                                                            
\centerline{\epsfig{file=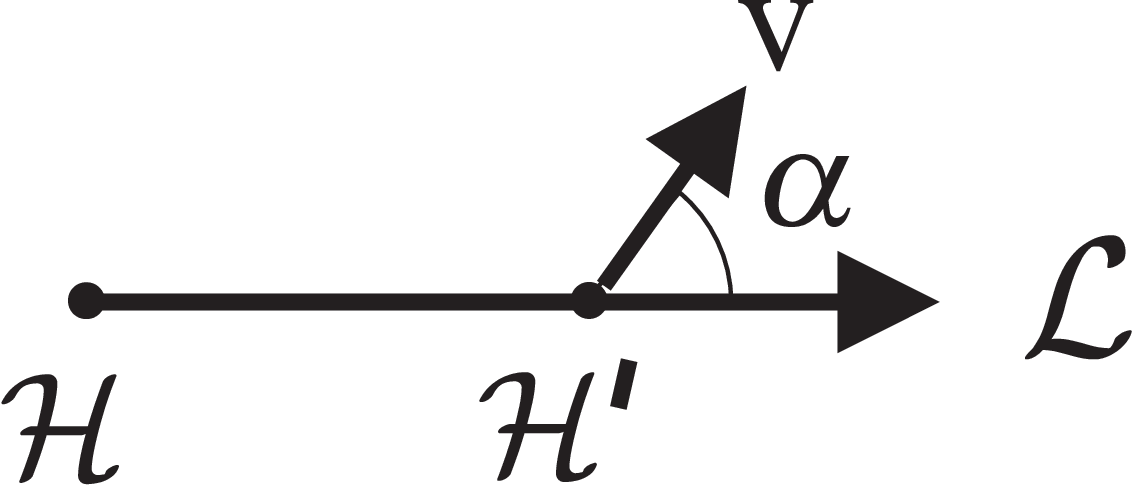, width=34mm, height=14mm}}
\selectlanguage{esperanto}\caption{\selectlanguage{esperanto}
La horlo\^goj $\cal H$ (mia, ripozanta) kaj $\cal H'$ (movi\^ganta).
\ppdu{\newline \selectlanguage{english}Figure \ref{FDoppler}:
The clocks $\cal H$ (mine, at rest) and $\cal H'$ (in motion).}}
\label{FDoppler}
\end{figure}

\ppsubsection{\label{transv}Transverso, $\alpha=\pi/2$}{Transverse, $\alpha=\pi/2$}

\ppln{Se $\alpha=\pi/2$, la horlo\^go $\cal H'$ momente movi\^gas flanke rilate al mi, do la distanco inter ni estas momente konstanta. Mi vidas, ke dum la montro de mia $\cal H$ pligrandi\^gas $\dd\tau_p$, la montro de $\cal H'$ pligrandi\^gis apena\u u~\cite[pa\^go~118]{LL}}
\pprn{If $\alpha=\pi/2$, the clock $\cal H'$ momentarily moves sideways relative to me, so the distance between us is momentarily constant.  I see that while the reading of my $\cal H$ increases $\dd\tau_p$, the reading of $\cal H'$ has increased only~\cite[page~118]{LL}}

\bea                                                                              \label{tempo2}
\dd\tau'=\sqrt{1-v^2/c^2}\,\dd\tau_p.
\eea

\ppln{Tiu $\!$ malsamo $\!$ de $\!$ montroj $\!$ nomi\^gas $\!$ transversa $\!$ Dopplera efiko. \^Gi anka\u u nomi\^gas alru\^go, \^car ordinara lumo irigita el $\cal H'$ \^sajnas delokita al ru\^go, kiam vidata per mi. La nerelativeca anta\u uvido estas $\dd\tau'=\dd\tau_p$\,, tial la transversa Dopplera efiko estas dirita nemikse relativeca.}
\pprn{That difference of readings is called transverse Doppler effect. It is also called redshift, because ordinary light emitted from $\cal H'$ appears shifted towards the red, as seen by me. The nonrelativistic prediction is $\dd\tau'=\dd\tau_p$\,, so the transverse Doppler effect is said purely relativistic.}

\ppsubsection{\label{movafast}Fori\^go, $\alpha=0$}{Removal, $\alpha=0$}

\ppln{Se $\alpha=0$, la horlo\^go $\cal H'$ radiuse fori\^gas de mi. Dum la montro de mia $\cal H$ pligrandi\^gas $\dd\tau_p$, mi vidas ke la montro de $\cal H'$ pligrandi\^gis apena\u u \cite[pa\^go~117]{LL}}
\pprn{If $\alpha=0$, the clock ${\cal H'}$ is radially moving away from me. While the reading of my $\cal H$ increases  $\dd\tau_p$, I see that the reading of $\cal H'$ has increased only \cite[page~117]{LL}}

\bea                                                                               \label{afast}
\dd\tau'=\frac{\sqrt{1-v^2/c^2}}{1+v/c}\dd\tau_p
=\sqrt{\frac{1-v/c}{1+v/c}}\dd\tau_p.
\eea

\ppln{Ni rimarkas, ke la relativeca Dopplera efiko \^ce fori\^go estas pli forta ol tiu \^ce transversa movado (\ref{tempo2}), kun sama rapido $v$\,. Nerelativece, $\dd\tau'=\dd\tau_p/(1+v/c)$. \^Car en amba\u u teorioj, relativeca kaj nerelativeca, veri\^gas $\dd\tau'<\dd\tau_p$\,, tial amba\u u anta\u uvidas alru\^gon. Rimarku, ke la relativeca variado $\dd\tau_p-\dd\tau'$ de fori\^go (\ref{afast}) estas pli granda ol la nerelativeca.}
\pprn{We remark that the relativistic Doppler effect of removal is stronger than that of transverse motion (\ref{tempo2}), with same speed $v$\,. Nonrelativistically, $\dd\tau'=\dd\tau_p/(1+v/c)$. Since in both theories, relativistic and nonrelativistic, it happens $\dd\tau'<\dd\tau_p$\,, both predict redshift. Remark that the relativistic shift $\dd\tau_p-\dd\tau'$ of removal (\ref{afast}) is larger than the nonrelativistic one.}

\ppsubsection{Alproksimi\^go, $\alpha=\pi$}{Approaching, $\alpha=\pi$}

\ppln{Se $\alpha=\pi$, la horlo\^go $\cal H'$ radiuse proksimi\^gas al mi. Nun mi vidas la montron de $\cal H'$ pligrandi\^gi pli rapide ol tiun de mia $\cal H$. Dum la montro de mia $\cal H$ pligrandi\^gas $\dd\tau_p$, mi vidas ke la montro de $\cal H'$ pligrandi\^gis \cite[pa\^go~117]{LL},}
\pprn{If $\alpha=\pi$, the clock $\cal H'$ is radially approaching me. Now I see the reading of $\cal H'$ increase faster than that of my $\cal H$. While the reading of my $\cal H$ increases $\dd\tau_p$, I see that the reading of $\cal H'$ has increased \cite[page~117]{LL},}

\bea                                                                               \label{aprox}
\dd\tau'=\frac{\sqrt{1-v^2/c^2}}{1-v/c}\,\dd\tau_p
=\sqrt{\frac{1+v/c}{1-v/c}}\,\dd\tau_p.
\eea

\ppln{Nerelativece $\dd\tau'=\dd\tau_p/(1-v/c)$. Tiu efiko nomi\^gas Dopplera de radiusa alproksimi\^go, a\u u albluo. Kontra\u ue okazo~\ref{movafast} (fori\^go), la relativeca albluo $\dd\tau'-\dd\tau_p$ estas malpli forta ol la nerelativeca.}
\pprn{Nonrelativistically $\dd\tau'=\dd\tau_p/(1-v/c)$. That effect is called Doppler of radial approach, or blueshift. Oppositely to the case~\ref{movafast} (removal), the relativistic blueshift $\dd\tau'-\dd\tau_p$ is weaker than the nonrelativistic.}

\ppl{Pasante, rimarku  en~(\ref{aprox}) la {\em faktoron k} de Bondi \cite[pa\^go~387]{Bondi}, $\sqrt{(1+v/c)/(1-v/c)}$.}
\ppr{By the way, remark in~(\ref{aprox}) the {\em k factor} of Bondi \linebreak[3]\cite[page~387]{Bondi}, $\sqrt{(1+v/c)/(1-v/c)}$.}

\ppsubsection{\label{casogeral}\^Generalo, $\alpha=\rm iom\; ajn$}{General, $\alpha=\rm any$} 

\ppln{\^Ci tiu okazo kunigas la 3 anta\u uajn okazojn, kaj ilin \^generaligas. Dum la montro de mia $\cal H$ pligrandi\^gas ${\rm d}\tau_p$, mi vidas ke tiu de $\cal H'$ pligrandi\^gis~\cite[pa\^go~118]{LL}}
\pprn{This case encompasses the 3 preceding ones, and generalizes them. While the reading of my $\cal H$ increases ${\rm d}\tau_p$, I see that the reading of $\cal H'$ has increased~\cite[page~118]{LL}}

\bea                                                                               \label{geral}
\dd\tau'=\frac{\sqrt{1-v^2/c^2}}{1+(v/c)\cos\alpha}\dd\tau_p.
\eea

\ppln{Nerelativece $\dd\tau'=\dd\tau_p/[1+(v/c)\cos\alpha]$\,. Simpla kalkulo $\!$ montras $\!$ ke $\!$ se $\!$ $0\leq|\alpha|<\pi/2$\,, $\!$ la nerelativeca kaj la relativeca anta\u uvidoj amba\u u estas alru\^go ($\dd\tau'<\dd\tau_p$). Sed se $\pi/2<|\alpha|\leq\pi$, la analizo estas malpli simpla: Newtone \^ciam okazas albluo ($\dd\tau'>\dd\tau_p$), kontra\u ue relativece okazas anka\u u albluo se nur la rapido de $\cal H'$ estas sufi\^ce malgranda, $v/c<2/|\cos\alpha+\sec\alpha|$; se tamen la rapido estas pli granda, la relativeca teorio anta\u uvidas alru\^gon ($\dd\tau'<\dd\tau_p$), kvankam la distanco de $\cal H'$ al $\cal H$ plieti\^gas.}
\pprn{Nonrelativistically $\dd\tau'=\dd\tau_p/[1+(v/c)\cos\alpha]$\,. A simple calculus shows that if $0\leq|\alpha|<\pi/2$, the nonrelativistic and the relativistic predictions both are redshift ($\dd\tau'<\dd\tau_p$). But if $\pi/2<|\alpha|\leq\pi$, the analysis is less simple: Newtonianly it always happens blueshift ($\dd\tau'>\dd\tau_p$), while relativistically it also happens blueshift only if the speed of $\cal H'$ is low enough, $v/c<2/|\cos\alpha+\sec\alpha|$; if however the speed is higher, relativity predicts redshift ($\dd\tau'<\dd\tau_p$), even though the distance between $\cal H'$ and $\cal H$ is decreasing.}

\ppsubsection{Logaritma spiralo}{Logarithmic spiral}

\ppln{Ni rimarkas ke en~(\ref{tempo2}), en kiu $\alpha=\pi/2$, veri\^gas $\dd\tau'<\dd\tau_p$, kaj ke en~(\ref{aprox}), en kiu $\alpha=\pi$, veri\^gas $\dd\tau'>\dd\tau_p$. Do, trudante $\dd\tau'=\dd\tau_p$ en~(\ref{geral}), ni trovas la mezan valoron de $\alpha$ en la intervalo ($\pi/2, \pi$) kiu momente nuligas la relativecan {\mbox Doppleran} efikon:} 
\pprn{We remark that in~(\ref{tempo2}), where $\alpha=\pi/2$, it happens $\dd\tau'<\dd\tau_p$, and that in~(\ref{aprox}), where $\alpha=\pi$, it happens $\dd\tau'>\dd\tau_p$. So, setting $\dd\tau'=\dd\tau_p$ in~(\ref{geral}), we find the intermediate value of $\alpha$ in the range ($\pi/2, \pi$) that momentarily makes zero the relativistic Doppler effect:}

\bea                                                                             \label{qqcoisa}
\tan^2(\alpha/2)=\sqrt{\frac{1+v/c}{1-v/c}}.
\eea 

\ppln{Tiu $\alpha$ pendas nur de la rilato $v/c$\,, kaj estas plikompakte skribita kiel}
\pprn{That $\alpha$ depends only on the ratio $v/c$\,, and is written more compactly as}

\bea                                                                              \label{logar2}
\cos\alpha=-\tanh(\xi/2), \hspace{3mm} \tanh\xi:=v/c.
\eea

\ppl{Por da\u ure esti $\dd\tau'=\dd\tau_p$ kun $v$ konstanta, $\alpha$ bezonas esti konstanta. En tiu okazo, estante $\cal H$ en la origino $r=0$, la trajektorio de $\cal H'$ estas konver\^ga logaritma spiralo:}
\ppr{To be permanently $\dd\tau'=\dd\tau_p$ with $v$ cons\-tant, $\alpha$ needs be constant. In this case, if $\cal H$ is at the origin  $r=0$, the trajectory of $\cal H'$ is a convergent logarithmic spiral:}

\bea                                                                              \label{logar3}
r=r_0\,\exp[-|\varphi|\,\sinh(\xi/2)].
\eea

\ppln{Tiu \^ci rezulto kontrastas kun tiu de nerelativeca teorio, kiu diras, ke la Dopplera efiko nuli\^gas, $\dd\tau'=\dd\tau_p$\,, nur se la luma fonto movi\^gas cirkle, $\alpha=\pm\pi/2$\,.}
\pprn{This result contrasts with that of the nonrelativistic theory, which says that the Doppler effect is null, $\dd\tau'=\dd\tau_p$\,, only if the light source has circular motion, $\alpha=\pm\pi/2$\,.}

\ppsection{\label{gemeos}\^Gemel-paradokso}{Twin paradox}

\ppln{Multe klarigan analizon de la horlo\^g-para\-dokso (a\u u \^gemel-) faris Darwin~\cite{Darwin}, uzante la esprimojn~(\ref{afast}) kaj~(\ref{aprox}). Esti\^gu komence apudaj, du ripozantaj horlo\^goj $\cal H$ kaj $\cal H'$, kaj esti\^gu punkto $P$ fiksita 4 lumjaroj for.}
\pprn{A very clarifying analysis of the clock (or twin) paradox was made by Darwin~\cite{Darwin}, using the expressions~(\ref{afast}) and~(\ref{aprox}). Let two clocks $\cal H$ and $\cal H'$ be initially close and at rest, and let a point $P$ be fixed 4 light-years away.}

\ppl{En iu momento, $\cal H'$ ekforiras direkte al $P$, kun konstanta rapido $v=4c/5$, dum $\cal H$ da\u ure ripozas. Dum la foriro, \^ciu horlo\^go vidas (nudokule a\u u dulornete a\u u teleskope) ke la takton de la alio estas pli malrapide ol la sian, \^ce la relativeca rilato $\sqrt{1-v/c} \div \sqrt{1+v/c} = \cdots =1/3$, kiel en ek.~(\ref{afast}); t.e., horlo\^go kies montro pligrandi\^gas 1 horon vidas, ke la montro de la alio pligrandi\^gis apena\u u 20 minutojn.}
\ppr{In some moment, $\cal H'$ departs towards $P$, at constant speed $v=4c/5$, while $\cal H$ remains at rest. During the removal, each clock sees (with naked eyes, or through a binocular, or through a telescope) the pace of the other to be slower then his own, at the relativistic ratio $\sqrt{1-v/c} \div \sqrt{1+v/c} = \cdots =1/3$, as in eq.~(\ref{afast}); that is, a clock whose reading increases 1 hour sees that the other has increased only 20 minutes.}

\ppl{\^Gisirante $P$, $\cal H'$ tuj revenas direkte al $\cal H$, kun rapido $4c/5$ denove. Kaj {\em poste iom da tempo}, dum la alproksimi\^go, \^ciu horlo\^go vidas, ke la montro de la alio pligrandi\^gis pli rapide ol la sia, \^ce la relativeca rilato $\sqrt{1+v/c} \div \sqrt{1-v/c} = \cdots =3/1$, kiel en ek.~(\ref{aprox}); t.e., horlo\^go kies montro pligrandi\^gas 1 horon vidas, ke tiu de la alio pligrandi\^gis 3 horojn.}
\ppr{Arriving at $P$, $\cal H'$ immediately comes back towards $\cal H$, with speed $4c/5$ again. And {\em after a while}, in the reapproximation, each clock sees that the reading of the other has increased more rapidly than his reading, in the relativistic ratio $\sqrt{1+v/c} \div \sqrt{1-v/c} = \cdots =3/1$, as in eq.~(\ref{aprox}); that is, a clock whose reading increases 1 hour sees that the other increased 3 hours.}

\ppl{\^Cio \^sajnas simetria, do la du horlo\^goj \^sajne montrus la saman registron en la renkonto; sed ni sekvu la pli detalajn malkunajn analizojn de $\cal H'$ kaj $\cal H$.}
\ppr{Everything seems symmetric, so the two clocks apparently should show the same reading at the re-encounter; but let us follow the more detailed reasonings of $\cal H'$ and $\cal H$, separately.}

\ppl{Jen la rezonado de la movi\^ganta horlo\^go $\cal H'$\,: mi ekforiras \^ce voja\^go, kies forira da\u uro estas sama kiel la revena, amba\u u mezuritaj per mi. Dum la unua duono de mia voja\^go mi vidos la montron de $\cal H$ pligrandi\^gi pli malrapide ol mian, \^ce la rilato $1/3$; kaj dum la dua duono mi vidos la montron de $\cal H$ pligrandi\^gi pli rapide ol mian, \^ce la rilato $3/1$. Do, meznombre, mi konstatos ke la takto de $\cal H$ estis  $(1/2)[(1/3)+(3/1)]=5/3$ de mia.}
\ppr{Here is the reasoning of the moving clock $\cal H'$\,: I am departing to a trip, whose duration in the removal is equal to the duration in the reapproximation, both measured by me. In the first half of my trip I will see the reading of $\cal H$ increase slower than mine, at the ratio $1/3$; and in the second half I will see the reading of $\cal H$ increase faster than mine, at the ratio 3/1. So, in the mean, I will find that the pace of $\cal H$ was $(1/2)[(1/3)+(3/1)]=5/3$ of mine.}

\ppl{La restanta horlo\^go $\cal H$ same konkludas per la jena analizo: $\cal H'$ ekforiris \^ce voja\^go de 5 foriraj jaroj kaj 5 revenaj jaroj, \^ciuj mezuritaj per mi; en nia renkonto mi certe montros $\Delta\tau=10$ jarojn. Dum iom da tempo mi vidos la montron de $\cal H'$ pligrandi\^gi pli malrapide ol mian montron, \^ce la rilato $1/3$, poste mi vidos \^gin pligrandi\^gi pli rapide ol mian, \^ce la inversa rilato $3/1$. Klare $\cal H'$ atingos $P$ kiam mi montros $\Delta\tau=5$ jarojn. \^Car $P$ distancas 4 lumjarojn de \^ci tie, tial la informo pri la ekreveno de $\cal H'$ atingos min nur en 4 jaroj poste. Do mi vidos la \^san\^gon de rilato 1/3 al rilato 3/1 nur kiam mi montros $\Delta\tau=9$ jarojn, kaj dum apena\u u la 1 jaro restanta mi vidos la montron de $\cal H'$ pligrandi\^gi pli rapide ol mian, \^ce la rilato 3/1. Meznombre mi konstatos, ke la takto de $\cal H'$ estas $(9/10)(1/3)+(1/10)(3/1)=3/5$ de mia; do en la renkonto, \^car mi montros $\Delta\tau=10$ jarojn, tial $\cal H'$ montros nur $\Delta\tau'=6$ jarojn.}
\ppr{The resting clock $\cal H$ concludes the same, through the following analysis: $\cal H'$ departed to a trip, 5 years going away and 5 years coming back, all measured by me; in our re-encounter I will certainly show $\Delta\tau=10$ years. For some time I will see the reading of $\cal H'$ increase more slowly than mine, at the ratio 1/3, later I will see it increase faster than mine, at the inverse ratio 3/1. Clearly $\cal H'$ will arrive at $P$ when I will show $\Delta\tau=5$ years. Since $P$ is 4 light-years far from here, the inform of the reverse motion of $\cal H'$ will reach me only 4 years later. So I will see the change of ratio 1/3 to ratio 3/1 only when I will show $\Delta\tau=9$ years, and only during the remaining 1 year I will see the reading of  $\cal H'$ increase faster than mine, at the ratio 3/1. In the mean, I will find that the pace of $\cal H'$ was $(9/10)(1/3)+(1/10)(3/1)=3/5$ from mine; so in the re-encounter, since I will show $\Delta\tau=10$ years, $\cal H'$ will show $\Delta\tau'=6$ years only.}

\ppl{\^Car tiu rilato 3/5 per $\cal H$ estas inversa de la rilato 5/3 per $\cal H'$, tial la du horlo\^goj same konkludas: $\cal H$, kiu \^ciam estis senmova en la sama inercia sistemo de referenco, montros ``pli sekundojn'' ol la voja\^ganta $\cal H'$, en la renkonto.}
\ppr{Since the ratio 3/5 found by $\cal H$ is the inverse of the ratio  5/3 found by $\cal H'$, the two clocks conclude the same: $\cal H$, who was always at rest in the same inertial reference system, will show ``more seconds'' than the travelling $\cal H'$, at the re-encounter.}

\ppsection{\label{sinc}Sinkroneco la\u u Einstein}{Einstein synchronism}

\ppln{La montro $\tau$ de horlo\^go ofte ne gravas, por ordinaraj analizoj; vere, ni plej ofte bezonas scii apena\u u intervalojn $\Delta\tau$. Nun supozu du horlo\^gojn, amba\u u ripozantaj en la sama inercia sistemo de referenco~\cite[pa\^go~1]{LL}, do havante la saman takton. Sed dum iu montras $\tau$, la alio povas montri $\tau^\star\neq\tau$. Tiam ni diras ke ili estas nesinkronaj, kaj ni volas sinkronigi ilin.} 
\pprn{The reading $\tau$ of a clock often does not {\mbox matter}, in ordinary analyses; indeed, we most often need know only intervals $\Delta\tau$. Now suppose two clocks, both at rest in the same inertial reference system~\cite[page~1]{LL}, so having the same pace. But while one shows $\tau$, the other may show $\tau^\star\neq\tau$. We then say they are out of synchronism, and we want to synchronize them.}

\ppl{Einstein sugestis la jenan recepton (difinon) por sinkronigi ripozantajn horlo\^gojn, $\cal H$ kaj $\cal H^\star$: en sia momento $\tau_1$\,, $\cal H$ eligas lumsignalon en la direkto al $\cal H^\star$; kiam la signalo atingas $\cal H^\star$, \^gi estas tuj reflektita revene al $\cal H$\,, kiu montras $\tau_2$ kiam la signalo alvenas. Tiam la kvanto $\Delta\tau:= (\tau_2-\tau_1)/2$ estas kalkulita. En iu ajn posta momento $\tau$ de $\cal H$, nova signalo estas eligita de $\cal H$ al $\cal H^\star$, kun la instruo ke, kiam la signalo atingas $\cal H^\star$, la montro de $\cal H^\star$ estu tuj \^san\^gata al $\tau+\Delta\tau$. Se la instruo estas obeita, la du horlo\^goj $\cal H$ kaj $\cal H^\star$ estas sinkronaj, kaj tiel ili restos.}
\ppr{Einstein suggested the following prescription (definition) to synchronize clocks at rest, $\cal H$ and $\cal H^\star$: in its moment $\tau_1$\,, $\cal H$ emits a light signal towards $\cal H^\star$; when the signal reaches $\cal H^\star$, it is immediately reflected back to $\cal H$\,, that shows $\tau_2$ when the signal arrives. Then the quantity $\Delta\tau:= (\tau_2-\tau_1)/2$ is computed. At any later moment $\tau$ of $\cal H$, a new signal is emitted from $\cal H$ towards $\cal H^\star$, with the prescription that, when the signal reaches $\cal H^\star$, the reading of $\cal H^\star$ be immediately changed to $\tau+\Delta\tau$. If the prescription is obeyed, the clocks $\cal H$ and $\cal H^\star$ are synchronized, and so they will remain.}

\ppsection{\label{Srekto}Rekta vico de horlo\^goj}{Straight row of clocks}

\ppln{Sinkronaj horlo\^goj restas la\u ulonge rekta segmento. Por ilin vidi, mi loki\^gas tre malproksime de la segmento, en direkto normala al \^gi. Plue, mi anka\u u havas \^ce mi horlo\^gon.}
\pprn{Synchronized clocks are disposed at rest along a straight segment. To see them, I stay very far from the segment and in direction normal to it. Further, I also have a clock with me.}

\ppl{En iu momento, la horlo\^ga vico ekmovi\^gas paralele al si, kaj tuj atingas konstantan rapidon $v$. \^Ciuj horlo\^goj sam-momente ekiris, la\u u atestata per siaj montroj, kaj konfirmita per mi en mia loko. Mi anka\u u vidas ke \^ciuj same akceli\^gis \^gis atingi la konstantan rapidon $v$. Simile kiel en la okazo~\ref{transv}, mi vidas ke la takto de \^ciu movi\^ganta horlo\^go estas {\em malpli rapida} ol mia, \^ce la rilato $\sqrt{1-v^2/c^2}$; kaj mi anka\u u vidas, ke ili da\u uras sinkronaj.}
\ppr{In some moment, the row of clocks starts moving parallel to itself, and soon reaches {\mbox constant} speed $v$\,. All clocks departed at the same moment, according to their registers, and confirmed by me from my position. I also see that they all had same acceleration till reach the constant speed $v$. Similarly as in case~\ref{transv}, I see that the pace of every clock in motion is {\em slower} than mine, in the ratio  $\sqrt{1-v^2/c^2}$; and I also see that they remain synchronous.}

\ppl{Anka\u u la movi\^gantaj horlo\^goj kun konstanta rapido $v$ vidas, ke iliaj taktoj estas la sama; sed ili vidas, ke ilia takto estas {\em pli rapida} ol la mia, \^ce tiu rilato $\sqrt{1-v^2/c^2}$. Tamen ili rimarkas, ke ili nun estas nesinkronaj. Ili rimarkas, ke la frontaj horlo\^goj de la movi\^ganta vico montras {\em pli} sekundojn ol la malfrontaj. Por distanco $l$, mezurita per ili, la malsinkrono estas $lv/c^2$ (\cite[pa\^go~11]{LL}, a\u u vidu  sekcion~\ref{rig} de tiu \^ci artiklo).}
\ppr{Also the clocks in motion with constant speed $v$ see that their paces are equal; but they see that their pace is {\em faster} than mine, in that ratio $\sqrt{1-v^2/c^2}$. But they remark that they are now out of synchronism. They remark that the front clocks of the moving row show {\em more} seconds than those on the rear. For a separation $l$, measured by them, the dis-synchronism is $lv/c^2$ (\cite[page~11]{LL}, or see section~\ref{rig} of this article).}

\ppl{Nun supozu, ke la horlo\^goj resinkroni\^gas dum la movado kun konstanta rapido. {\mbox Poste} la resinkroni\^go, mi ilin vidus nesinkronaj, tiuj frontaj montrante {\em malpli} sekundojn ol tiuj malfrontaj. Por distanco $l$, mezurita per la horlo\^goj, la malsinkrono vidata per mi estus $lv/c^2$.}
\ppr{Now suppose the clocks re-syn\-chro\-ni\-ze, in the motion with constant velocity. After re-synchronization, I would see them out of synchronism, the front clocks now showing {\em less} seconds than those of the rear. For a separation $l$, measured by the clocks, the dis-synchronism seen by me would be $lv/c^2$.}

\ppl{En sekcioj~\ref{rig} kaj~\ref{acelconst} ni diskutos alian kielon por akceli horlo\^gojn: ili estos fiksitaj en rigida stango.}
\ppr{In sections~\ref{rig} and~\ref{acelconst} we discuss another way of accelerating clocks: they will be fixed in a rigid rod.}

\ppsection{\label{circulo}Cirklo de horlo\^goj}{Circle of clocks}

\ppln{Imagu vicon de sinkronaj horlo\^goj, fiksitaj en cirklo kun radiuso $R$\,. La cirklo komence restas en inercia sistemo de referenco. Poste, \^gi ekrotacias \^cirka\u u sia orta akso, \^gis atingi {\mbox konstantan} angulan rapidon $\omega$\,. \^Ciuj horlo\^goj sam-momente ekmovi\^gis, la\u u atestata per siaj montroj, kaj \^ciuj angule akceli\^gis same.}
\pprn{Imagine a string of synchronized clocks, fixed on a circle with radius $R$\,. The circle is initially at rest in an inertial system of reference. Then it starts rotating around its normal axis, {\mbox until} reach constant angular speed $\omega$. All clocks started moving in the same moment, according to their registers, and all angularly accelerated the same.}

\ppl{Mi estas fiksita en cirklocentro. Mi konfirmas ke mi vidis \^ciujn horlo\^gojn ekiri samtempe, kaj mi vidas ke iliaj taktoj fari\^gis pli malrapide ol tiu de mia horlo\^go (okazo~\ref{transv}), \^ce la rilato $\sqrt{1-R^2\omega^2/c^2}$. Do la periodo de la cirkla movado, kiel mezurata per la movi\^gantaj horlo\^goj, estas malpli granda ol tiu mezurata per mi, je tiu rilato. Mi rimarkas anka\u u, ke ilia sinkrono ne malfari\^gis, simile kiel en la okazo de rekta movado.}
\ppr{I am fixed at the centre of the circle. I confirm that I saw all clocks started moving simultaneously, and I see that their pace became slower than mine (case~\ref{transv}), in the ratio $\sqrt{1-R^2\omega^2/c^2}$. So the period of the circular motion, as measured by the clocks in motion, is less than that measured by me, in that ratio. I also remark that their synchronism was maintained, similarly as in the case of rectilinear motion.}

\ppl{\^Ciu horlo\^go en la rotacianta cirklo rimarkas, ke la alioj ankora\u u havas takton same kiel sian, sed \^gi rimarkas, ke \^gi perdis la sinkronon kun ili. \^Gi rimarkas, ke la horlo\^goj apena\u u frontaj en la movado, montras pli sekundojn ol si, kontra\u ue, tiuj ke estas apena\u u malfrontaj montras malpli sekundojn. \^Gis nun nia raporto koincidas kun tiu de la rekta movado.} 
\ppr{Each clock of the rotating circle remarks that the others still have pace equal to his, but remarks that it lost the synchronism with them. It remarks that the clocks just before him, in the motion, show more seconds than him, while those just behind show less seconds. Till this point our description coincides with that of the rectilinear motion.}

\ppl{Ni vidu nun la malsamojn; ili okazas \^car la horlo\^goj estas fikitsaj en neinercia sistemo de referenco, do la relativeca simetrio estas rompita. \^Ciu horlo\^go de la rotacianta cirklo vidas, ke sia takto estas pli {\em malrapida} ol la mia, {\mbox haltadita} en la cirklocentro. Plue ne estas ebla, ke la horlo\^goj en cirkla movado {\it \^ciue} sinkroni\^gas la\u u Einstein.}
\ppr{Now let us see the differences; these {\mbox occur} because the clocks are now fixed on a non-inertial reference system, so the relativistic symmetry is broken. Each clock on the {\mbox rotating} circle sees that his pace is {\em slower} than mine, staying at the centre of the circle. Besides, it is impossible that {\it all} the clocks in circular motion re-synchronize with the Einstein prescription.}

\ppsection{\label{periodo}Periodo}{Period} 

\ppln{Kiel ekzerco, ni nun rederivos la rilaton, inter la du malsimilaj periodoj por la cirkla movado, findita en anta\u ua sekcio. Sed nun, mi kaj mia horlo\^go $\cal H$ estas haltaditaj en fiksita punkto apud la rotacianta cirklo. Do ni uzos la ekvacion~(\ref{geral}) pri la \^generala Dopplera efiko, en kiu, kvankam la modulo $v=R\omega$ de la rapido de horlo\^go $\cal H'$ estas konstanta, la angulo $\alpha$ ne estas; vidu figuron~\ref{Ftpost}.}
\pprn{As an exercise, we now reobtain the relation, between the two different periods for the circular motion, found in the preceding section. But now, I and my clock $\cal H$ are at rest in a fixed point close to the rotating circle. So we use the equation~(\ref{geral}) of the general Doppler effect, where although the modulus $v=R\omega$ of the speed of clock $\cal H'$ is constant, the angle $\alpha$ is not; see the figure~\ref{Ftpost}.}

\ppl{Esti\^gu $\varphi_0=0$ la angula pozicio de amba\u u $\cal H$ kaj $\cal H'$ kiam $t=0$\,. Do la angula loko de $\cal H'$ en posta momento $t$ estas $\varphi(t)=2\alpha(t)=\omega t$, kaj la distanco inter la du horlo\^goj estas $l(t)=2\,R\,\sin\alpha(t) = 2\,R\,\sin(\omega t/2)$, kiel mezurita en inercia sistemo de referenco $S_0$ kie mi kaj $\cal H$ restas. La ek.~(\ref{geral}) esti\^gas}
\ppr{Let $\varphi_0=0$ be the angular position of both $\cal H$ and $\cal H'$ when $t=0$\,. So the angular position of $\cal H'$ in a later moment $t$ is  $\varphi(t)=2\alpha(t)=\omega t$, and the distance between the two clocks is $l(t)=2\,R\,\sin\alpha(t) = 2\,R\,\sin(\omega t/2)$, as measured in an inertial sistem of reference $S_0$ where I and $\cal H$ are at rest. The eq.~(\ref{geral}) becomes}


\bea                                                                            \label{taulinha}
\dd\tau'=
\frac{\sqrt{1-R^2\omega^2/c^2}}{1+(R\omega/c)\cos(\omega t/2)}\,\dd t_p.
\eea

\ppln{Mi komence ($t\approx0$) vidas fori\^gan Doppleran efikon kiu iom-post-iome esti\^gas transversa ($t=\pi/\omega$) kaj fine ($t\approx2\pi/\omega$) alproksimi\^ga. La rilato inter la periodoj estos donata per la integrado de la rilato~(\ref{taulinha}) je unu plena turno. Por tio konsideru, ke en momento $t$ de $S_0$\, lumsignalo ekiras de $\cal H'$ kaj atingas $\cal H$ en la posta momento}
\pprn{I initially ($t\approx0$) see Doppler effect of removal, which progressively becomes transverse ($t=\pi/\omega$) and finally ($t\approx2\pi/\omega$) of approximation. The relation between the periods is given by the integral of the relation~(\ref{taulinha}) in a complete cycle. To that end, consider that at an instant $t$ of $S_0$\, a light signal starts from $\cal H'$, and reaches $\cal H$ at the later moment}

\bea                                                                             \label{taupost}
t_p:=t+\frac{l(t)}{c}
=t+\frac{2R}{c}\sin(\omega t/2).
\eea

\ppln{Diferencialante \^ci tiun ekvacion ni ricevas}
\pprn{Differentiating this equation we obtain}

\bea                                                                           \label{taulinha2}
\dd t_p=[1+(R\omega/c)\cos(\omega t/2)]\,\dd t,
\eea

\ppln{kiu$\,$ simpligas$\,$ (\ref{taulinha})$\,$ al$\,$ $\dd\tau'=\sqrt{1-R^2\omega^2/c^2}\,\dd t$$\,$. Integrante$\,$ tiu$\,$ lasta$\,$ ekvacio$\,$ en unu turno, ni ricevas\, la\, rilaton\, inter\, la\, du\, periodoj\,: $\Delta\tau'=\sqrt{1-R^2\omega^2/c^2}\,(2\pi/\omega)$. Indas noti, ke integrante (\ref{taulinha2}) en unu turno, ni ricevas $\Delta t_p=2\pi/\omega$\,; tio estis esperinda \^car, post pleno de unu turno, la du horlo\^goj estas denove kunaj kaj do $\Delta t_p = \Delta t$.}
\pprn{which simplifies~(\ref{taulinha}) to $\dd\tau'=\sqrt{1-R^2\omega^2/c^2}\,\dd t$. Integrating this last equation in one turn, we obtain the relation between the two periods: $\Delta\tau'=\sqrt{1-R^2\omega^2/c^2}\,(2\pi/\omega)$. It is worth {\mbox noting} that, integrating (\ref{taulinha2}) in one turn, gives $\Delta t_p=2\pi/\omega$\,; this was expected, since after one turn the two clocks are again together and so $\Delta t_p=\Delta t$.}

\begin{figure}
\centerline{\epsfig{file=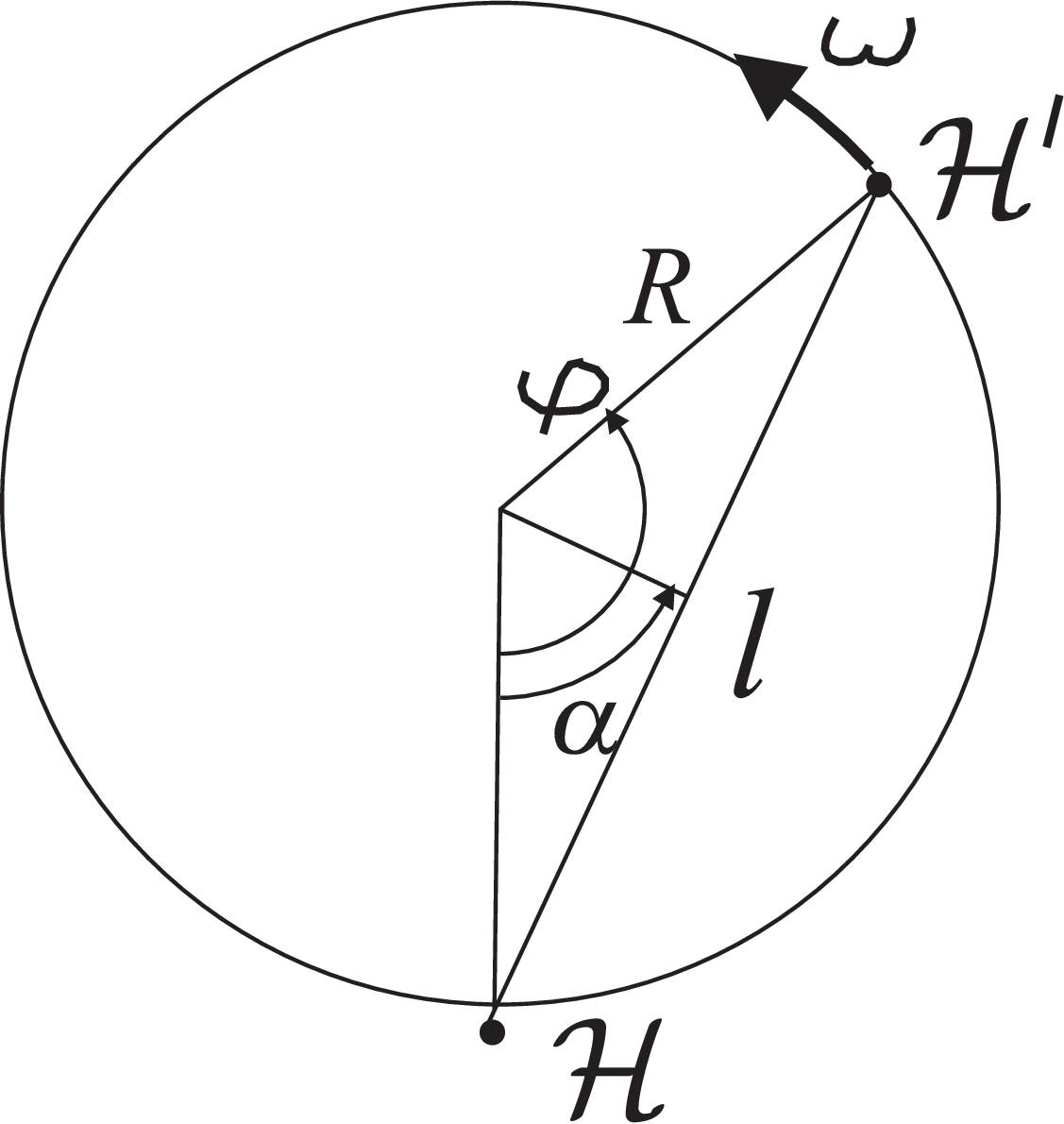,width=4cm,height=4cm}}
\selectlanguage{esperanto}\caption{\selectlanguage{esperanto}
La horlo\^go $\cal H$ ripozas, dume $\cal H'$ unuforme cirkle movi\^gas. 
\ppdu{\newline \selectlanguage{english}Figure \ref{Ftpost}:
The clock $\cal H$ is at rest, while $\cal H'$ has a uniform circular motion.}}
\label{Ftpost}
\end{figure}

\ppsection{\label{rig}Rigida stango}{Rigid rod}

\ppln{En sekcio~\ref{Srekto} ni movis etendeblan stangon same akcelante \^ciujn iliajn punktojn; tiu faris, ke la stanga propralongo pligrandi\^gu. Ni nun vidos kiel akceli tiujn punktojn, tiel ke la propralongo ne \^san\^gu.}
\pprn{In section~\ref{Srekto} we set in motion an extensible rod by equally accelerating all its points; that made increase the proper length of the rod. We now see how to accelerate these points, so that the proper length does not change.}    

\ppl{Esti\^gu inercia sistemo de referenco $S_0$, kun koordinatoj $x$ kaj $t$, kaj {\it rigida} stango komence ripozanta la\u ulonge de la akso $x$\,. Anka\u u esti\^gu punktoj $a$ kaj $b$ de la stango en la lokoj $0$ kaj $L$, respektive. En momento $t=0$\,, la stango ekmovi\^gas la\u ulonge de si, tiel ke $a$ movi\^gas la\u u $x(t)$. La stanga rigideco trudas ke $b$ movi\^gas tiel, ke la distanco $L$ inter $a$ kaj $b$ restas konstanta per \^ciu inercia sistemo de referenco $S_v$ kie $a$ momente ripozas.}
\ppr{Let be an inertial reference system $S_0$, with coordinates $x$ and $t$, and a {\it rigid} rod initially at rest along the $x$-axis. Also let points $a$ and $b$ of the rod at places $0$ and $L$, respectively. In moment $t=0$ the rod starts moving parallel to itself, with $a$ moving according to $x(t)$. The rigidity of the rod makes $b$ move in such a way that the separation $L$ between $a$ and $b$ remains constant in every inertial reference system $S_v$ in which $a$ is momentarily at rest.}


\ppl{Sur la figuro~\ref{Fbarrarigida} ni havas la eventojn $E=[x,ct]$, \^ce la historio de $a$, kaj $E_b=[x_b,ct_b]$, \^ce la historio de $b$\,; esti\^gu tiuj eventoj samtempaj en la inercia sistemo de referenco $S_v$. Tiuokaze, la angulo $\varphi$ rilatas al la rapido de $a$ la\u u  $\tan\varphi(t)=v(t)/c$, en kiu $v(t):=\dd x(t)/\dd t$. Ni volas rilatigi la valorojn de $x_b$ kaj $t_b$ al la valoroj de $x$ kaj $t$. Unue $\!$ $\!$ ni $\!$ vidas $\!$ sur $\!$ la $\!$ figuro, $\!$ ke $\!$ $\tan\varphi(t)=c(t_b-t)/(x_b-x)$, do}
\ppr{In figure~\ref{Fbarrarigida} we have the events $E=[x,ct]$, in the history of $a$, and $E_b=[x_b,ct_b]$, in the history of $b$\,; let these events be simultaneous in the inertial reference system $S_v$. In that case, the angle $\varphi$ is related to the speed of $a$ according to   $\tan\varphi(t)=v(t)/c$, in which $v(t):=\dd x(t)/\dd t$. We want to relate the values of $x_b$ and $t_b$ to the values of $x$ and $t$. First, we see on the figure that $\tan\varphi(t)=c(t_b-t)/(x_b-x)$, so}

\bea                                                                              \label{tempos}
c(t_b-t)=\frac{v(t)}{c}(x_b-x);
\eea

\ppln{poste ni trudas la Lorentzan konstantecon de la interspaco $L$, t.e.,}
\pprn{then we impose the Lorentzian constancy of the separation $L$, i.e.,}

\bea                                                                               \label{separ}
(x_b-x)^2-c^2(t_b-t)^2=L^2;
\eea

\ppln{fine ni ricevas de~(\ref{tempos}) kaj~(\ref{separ}) (Nikoli\'c~\cite{Nikolic})}
\pprn{finally we obtain from~(\ref{tempos}) and~(\ref{separ}) (Nikoli\'c~\cite{Nikolic})}

\bea                                                                                  \label{xb}
x_b = x+L\gamma, \hspace{5mm} t_b = t+\frac{L}{c^2}v\gamma, \hspace{5mm} \gamma:=\frac{1}{\sqrt{1-v^2/c^2}}\,.
\eea 

\ppln{Memoru, ke $x$ estas pozicio de punkto $a$ en {\mbox momento} $t$, dum ke $x_b$ estas pozicio de $b$ en posta momento $t_b$ per~(\ref{xb}b).}
\pprn{Remember that $x$ is the position of point $a$ at moment $t$, while $x_b$ is the position of $b$ at later moment $t_b$ through~(\ref{xb}b).}

\begin{figure}
\centerline{\epsfig{file=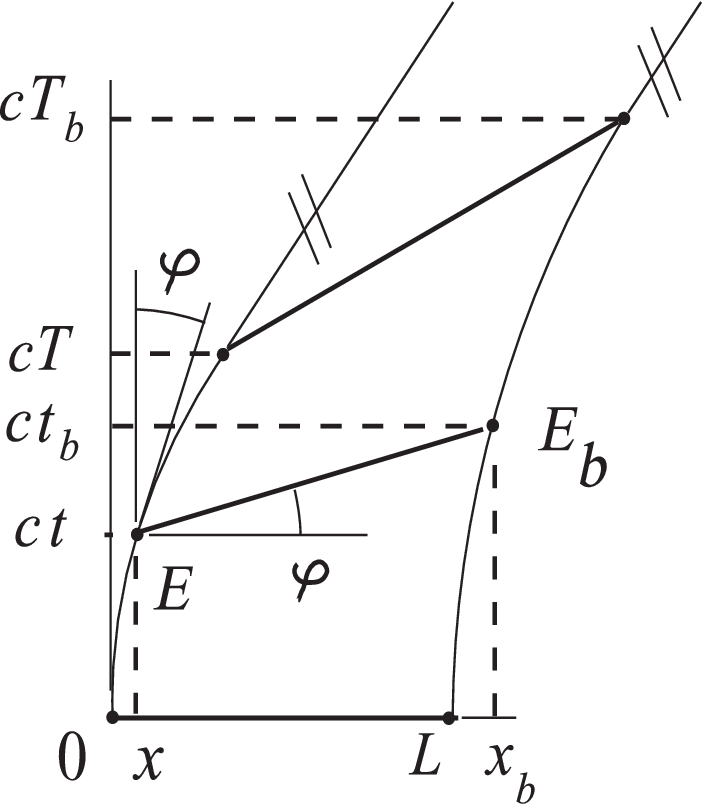,width=4cm}} 
\selectlanguage{esperanto}\caption{\selectlanguage{esperanto}
Grafika\^\j o de la historio de $a$, maldekstre, kaj $b$, dekstre. La punktoj $a$ kaj $b$ akceli\^gas \^gis la momentoj $T$ kaj $T_b$, respektive, poste iliaj rapidoj fari\^gas konstantaj kaj samaj. La eventoj $E$ kaj $E_b$ estas samtempaj per la inercia sistemo de referenco $S_v$, kies aksoj faras angulon $\varphi(t)$ kun la aksoj de $S_0$.
\ppdu{\newline \selectlanguage{english}Figure \ref{Fbarrarigida}:
Graphic of the history of $a$, at left, and $b$, at right. The points $a$ and $b$ are accelerated until the moments $T$ and $T_b$, respectively, then their speeds become constant and equal. The events $E$ and $E_b$ are simultaneous in the inertial reference system $S_v$, whose axes make angle $\varphi(t)$ with the axes of $S_0$.}}
\label{Fbarrarigida}
\end{figure}

\ppl{Ni montras, ke en \^ciuj sinsekvaj inerciaj {\mbox sistemoj} de referenco $S_v$, \^ciuj stangaj punktoj estas same rapidaj. \^Car punkto $b$ ne estas speciala, tiam sufi\^cas montri, ke \^gia rapido $\dd x_b/\dd t_b$ en momento $t_b$ estas $v(t)$. Por tio, ni unue {\mbox diferencialas} (\ref{xb}a) kaj (\ref{xb}b) kaj ricevas}
\ppr{We show that, in all successive inertial {\mbox reference} systems $S_v$, all points of the rod have same speed. Since the point $b$ has nothing {\mbox special}, it suffices to show that its speed $\dd x_b/\dd t_b$ in moment $t_b$ is $v(t)$. To that end, we first {\mbox differentiate} (\ref{xb}a) and (\ref{xb}b) and obtain}

\bea                                                                               \label{difxb}
\dd x_b &=& \dd x + (L/c^2)\gamma^3 v \dd v,
\\                                                                                 \label{diftb}
\dd t_b &=& \dd t+(L/c^2)\gamma^3 \dd v,
\eea

\ppln{en kiuj ni uzis}
\pprn{in which we used}

\bea                                                                                            
\dd\gamma\equiv(\gamma^3v/c^2)\,\dd v\,, \hspace{5mm} \dd(v\gamma)\equiv\gamma^3\,\dd v\,. 
\eea 

\ppln{Poste ni uzas~(\ref{difxb}) kaj~(\ref{diftb}) en $v_b(t_b)=\dd x_b/\dd t_b$, kaj ricevas efektive $v_b(t_b)=v(t)$ (kaj konsekvence $\gamma_b(t_b)=\gamma(t)$). Konvenas ripeti ke \^ci tiu samo de rapidoj de $a$ kaj $b$ okazas per la sinsekvaj inerciaj sistemoj de referenco $S_v$, kaj ne per la inercia sistemo de referenco $S_0$.}
\pprn{Then we use~(\ref{difxb}) and~(\ref{diftb}) in $v_b(t_b)=\dd x_b/\dd t_b$, and obtain effectively $v_b(t_b)=v(t)$ (and consequently $\gamma_b(t_b)=\gamma(t)$). It is worth repeating that this equality of speeds of $a$ and $b$ occurs in the successive inertial reference systems $S_v$, and not in the inertial reference system $S_0$.}

\ppl{Supozu, ke la punktoj $a$ kaj $b$ estas {\mbox sinkronaj} horlo\^goj en komenca momento $t=0$\,, kiam ili estis restantaj kaj montris $\tau=0$ kaj $\tau_b=0$\,. Nun ni montras, ke per la sinsekvaj inerciaj {\mbox sistemoj} de referenco $S_v$, ili ne restas sinkronaj, kaj ni kalkulas la mal-sinkronon. Por tio, ni {\mbox kalkulas} la propratempojn $\tau$ kaj $\tau_b$ montratajn per tiuj horlo\^goj ekde la ripozo. Integrante $\dd\tau=\dd t/\gamma(t)$ de 0 \^gis $t$, kaj $\dd\tau_b=\dd t_b/\gamma_b(t_b)$ de 0 \^gis $t_b$\,, ni ricevas}
\ppr{Suppose that the points $a$ and $b$ are synchronous clocks in the initial moment $t=0$\,, when they were at rest and showed $\tau=0$ and $\tau_b=0$\,. We now show that, in the successive inertial reference systems $S_v$, they do not remain synchronous, and calculate the dis-synchronism. To that end, we calculate the propertimes $\tau$ and $\tau_b$ shown by these clocks since the state of rest. Integrating $\dd\tau=\dd t/\gamma(t)$ from 0 to $t$, and $\dd\tau_b=\dd t_b/\gamma_b(t_b)$ from 0 to $t_b$\,, we obtain}

\bea                                                                                \label{taua}
\tau(t) &=& \int_0^t{\frac{\dd t}{\gamma(t)}},
\\ \nonumber 
\tau_b(t_b) &=& \int_0^{t_b}{\frac{\dd t_b}{\gamma_b(t_b)}}
=\int_0^t{\frac{\dd t}{\gamma(t)}}+\frac{L}{c^2}\int_0^{v} {\gamma^2\dd v}
=\tau(t) + \frac{L}{c^2}\int_0^{v}{\frac{\dd v}{1-v^2/c^2}}
\\                                                                                  \label{taub}
&=& \tau(t)+\frac{L}{c}\tanh^{-1}\frac{v(t)}{c};
\eea 

\ppln{en la supra kalkulo ni uzis~(\ref{diftb}) por \^san\^go de integra variablo, kaj rilataj limoj, de $t_b$ al $t$ kaj $v$. Ni vidas do, ke en \^ciu ripoza inercia sistemo de referenco $S_v$ de la rigida stango, la horlo\^goj montras malsinkronon}
\pprn{in this calculation we used~(\ref{diftb}) to change the variable of integration, and related limits, from $t_b$ to $t$ and $v$. We then see that, in every inertial reference system $S_v$ where the rigid rod is at rest, the clocks show a dis-synchronism}

\bea                                                                             \label{dessinc}
\tau_b(t_b)-\tau(t) = \frac{L}{c}\tanh^{-1}\frac{v(t)}{c},
\eea

\ppln{en kiu $L$ estas la propra interspaco inter la horlo\^goj. La malsinkrono~(\ref{dessinc}) estas fore vidata, per observanto  en vida rekto normala al la stango, kaj kun rapido $v$ en la inercia sistemo de referenco $S_0$. La rezulto~(\ref{dessinc}) \^generalas tion de Giannoni kaj Gr\o n~\cite{GiannoniGron}, kiuj havigis \^gin en la speciala okazo pri konstanta propra akcelo de $a$, la\u u ni vidos en la sekvanta sekcio.}
\pprn{where $L$ is the proper separation between the clocks. The dis-synchronism~(\ref{dessinc}) is seen, by an observer far away in a line of sight normal to the rod, and with speed $v$ in the inertial reference system $S_0$. The result~(\ref{dessinc}) generalizes that of Giannoni and Gr\o n~\cite{GiannoniGron}, who obtained it in the special case of constant proper acceleration of $a$, as we shall see in the next section.}


\ppl{Ni povas anka\u u montri, ke \^ciuj horlo\^goj fiksitaj en la rigida stango havas saman takton, per \^ciu inercia sistemo de referenco $S_v$. Por tio, ni kalkulas $\dd\tau=\dd t/\gamma$ pri la horlo\^go $a$ en momento $t$, kaj pri la horlo\^go $b$ en momento $t_b$ (rememoru ke la eventoj priskribitaj kiel $[x,ct]$ kaj $[x_b,ct_b]$ en $S_0$ estas samtempaj en $S_v$\,):}
\ppr{We can also show that all clocks fixed on the rigid rod have same pace, in every inertial reference system $S_v$. To that end, we calculate $\dd\tau=\dd t/\gamma$ for the clock $a$ at the moment $t$, and for the clock $b$ at the moment $t_b$ (remember that the events described as $[x,ct]$ and $[x_b,ct_b]$ in $S_0$ are simultaneoius in $S_v$\,):}

\bea                                                                              \label{detaua}
\dd\tau(t) = \frac{\dd t}{\gamma(t)} ,\hspace{4ex}
\dd\tau_b(t_b)=\frac{\dd t}{\gamma_b(t_b)};
\eea

\ppln{\^car $\gamma_b(t_b)=\gamma(t)$, tial ni notas ke efektive $\dd\tau_b(t_b)=\dd\tau(t)$. Konvenas emfazi ke tiu takta sameco en samaj momentoj estas vidata apena\u u per observanto en la sinsekvaj sistemoj de referenco $S_v$, kaj ne en la komenca sistemo de referenco $S_0$. Ja, en la inercia sistemo de referenco $S_0$\,, la horlo\^go $b$ pulsas en la momento $t_b$ same kiel la horlo\^go $a$ pulsis en la iam anta\u ua momento $t$\,.}
\pprn{since $\gamma_b(t_b)=\gamma$, we remark that effectively $\dd\tau_b(t_b)=\dd\tau(t)$. It is worth to emphasize that such equality of pace at same moments is seen only by an observer in the successive reference systems $S_v$, and not in the initial reference system $S_0$. Really, in the inertial reference system $S_0$ the clock $b$ beats in the moment $t_b$ as the clock $a$ beated in the preceding moment $t$\,.}

\ppl{\^Sajnas paradoksa ke \^ciuj horlo\^goj fiksitaj en la movi\^ganta rigida stango havas  {\em saman takton} per \^ciu sistemo de referenco $S_v$, kaj tamen iom-post-iome malsinkroni\^gas per la sinsekvaj $S_v$. La paradokso estas solvata se ni memoras ke kvankam \^ciu $S_v$ estas inercia, sistemo de referenco kiu akompanas la akcelan movadon de la rigida stango ne estas inercia. Notu ke la samtempa rekto $E$ -- $E_b$, sur figuro~\ref{Fbarrarigida}, \^san\^gas la klinon iom-post-iome.}
\ppr{It seems paradoxical that all clocks fixed on the moving rigid rod  have {\em same pace} in every reference system $S_v$, and nevertheless progressively dis-sinchronize in the successive $S_v$. The paradox is solved if we remember that while every $S_v$ is inertial, a reference system who accompanies the accelerated motion of the rigid rod is not inertial. Remark that the line of simultaneity $E$ -- $E_b$, in  figure~\ref{Fbarrarigida}, changes slope progressively.}

\ppl{Nun ni supozu, ke en momento $T$ la horlo\^go $a$ malda\u urigas sian akcelon, kaj tenas konstanta la rapidon $V:=v(T)$. Iom-post-iome la horlo\^goj anta\u uaj $a$ anka\u u havigos la konstantan rapidon $V$ \^gis kiam la horlo\^go $b$ anka\u u havos rapidon $V$, en la momento $T_b$. Do, en la momento $T_b$, \^ciu peco de rigida stango inter $a$ kaj $b$ havos la konstantan rapidon $V$ mezurata per la inercia sistemo de referenco $S_0$. Ni povas nun kalkuli la malsinkronon $\tau_b(T_b)-\tau(T_b)$ inter la horlo\^goj $a$ kaj $b$\,, en la momento $T_b$ de $S_0$\,: ni unue kalkulas}
\ppr{Let us now suppose that at moment $T$ the clock $a$ ceases its acceleration, and maintains constant the speed $V:=v(T)$. Progressively the clocks ahead $a$ in the motion will also have the constant speed $V$ until when the clock $b$ also have speed $V$, at moment $T_b$. So, in moment $T_b$, every piece of the rigid rod between $a$ and $b$ will have the constant speed $V$\,, as measured in the inertial reference system $S_0$. We can now calculate the dis-synchronism $\tau_b(T_b)-\tau(T_b)$ {\mbox between} the clocks $a$ and $b$\,, in the moment $T_b$ of $S_0$\,: we first calculate}

\bea                                                                               \label{tauab}
\tau(T_b)=\int_0^{T_{b}}{\frac{\dd t}{\gamma}}
= \left(\int_0^{T}+\int_{T}^{T_{b}}\right)\frac{\dd t}{\gamma}
= \tau(T)+\frac{T_b-T}{\gamma(T)}
= \tau(T)+\frac{LV}{c^2},
\eea                                                       

\ppln{en kiu, per~(\ref{xb}b), ni uzis $(T_b-T)/\gamma(T)=LV/c^2$. Konsidere~(\ref{dessinc}) kun $t_b=T_b$ kaj $t=T$, kaj~(\ref{tauab}), ni fine ricevas}
\pprn{where, through~(\ref{xb}b), we used $(T_b-T)/\gamma(T)=LV/c^2$. Considering~(\ref{dessinc}) with $t_b=T_b$ and $t=T$, and~(\ref{tauab}), we finally obtain}

\bea                                                                            \label{dessinc2}
\tau_b(T_b)-\tau(T_b) = \frac{L}{c}\tanh^{-1}(V/c)-\frac{LV}{c^2}.
\eea

\ppln{\^Ci tiu estas la malsinkrono per la inercia {\mbox sistemo} de referenco $S_0$ komence ripoza de la stango. Do observanto sufi\^ce malproksime de la stango, en direkto normala al \^gi (kiel en sekcio~\ref{Srekto}), vidas malsinkronon la\u u~(\ref{dessinc2}). \^Gi pendas nur de la valoro $V$, kaj ne de la {\mbox maniero} kiel tiu rapido estas ricevita. Notu ke se la horlo\^go $a$ da\u uri\^gas kun konstanta {\mbox rapido} post $T_b$, tiu malsinkrono da\u uri\^gas. \^Car 
$\tanh^{-1}(V/c)=V/c + (V/c)^3/3 + \cdots$, tial ni vidas ke la malsinkrono~(\ref{dessinc2}) inter la horlo\^goj estas \^ce ordo $LV^3/3c^4$, tre malgranda kaj malsigna ol la familiara $-LV/c^2$ trovata inter koordinataj horlo\^goj \^ce la Lorentzaj transformoj (lasta termo en (\ref{dessinc2})).}
\pprn{This is the dis-synchronism in the inertial {\mbox reference} system $S_0$ of the initial rest of the rod. So an observer sufficiently far from the rod, in a direction normal to it (as in section~\ref{Srekto}), sees a dis-synchronism as in~(\ref{dessinc2}). It depends only on the value $V$, and not on the way that such speed was obtained. {\mbox Remark} that if the clock $a$ maintains the constant speed {\mbox after} $T_b$, that dis-synchronism remains. Since  $\tanh^{-1}(V/c)=V/c+(V/c)^3/3+\cdots$, we see that the dis-synchronism (\ref{dessinc2}) between the clocks is of order  $LV^3/3c^4$, very small and of opposite sign of the familiar $-LV/c^2$ found between coordinate clocks in the Lorentzian transformations (the last term in (\ref{dessinc2})).}

\ppl{Indas noti, ke ni povas eviti la malsaman maljuni\^gon de la stang-punktoj, se ni ne trudas rigidecon dum la akcelo, t.e., se ni fiksas nur la kondi\^cojn komencajn (ripozon) kaj finajn (rapidon $V$). Ja, se iu stang-punkto zigzagas kun rapido proksima al $c$, \^gia propratempaj intervaloj estos tiel malgranda kiel ni volas.}
\ppr{It is worth noting that we can avoid the unequal aging of the points of the rod, if we do not impose rigidity during the acceleration, i.e., if we fix only the initial conditions (rest) and final (speed $V$). Indeed, if every point of the rod zigzags with speed near to $c$, its intervals of propertime will be as short as we want.}

\ppsection[0.6ex]{\label{acelconst}Konstanta \\ propra akcelo}{Constant \\ proper acceleration}

\ppln{En la anta\u ua sekcio ni studis la movadon de stango kiu restis rigida kun arbitra akcelo. Ni nun specialigas, elektante $x(t)$ tiel ke la horlo\^go $a$ havu konstantan propran akcelon
\cite[pa\^go~22]{LL}, \cite[pa\^go~73]{Moller}, \cite[pa\^go~49]{Rindler}, \cite{GiannoniGron} kaj \cite{Nikolic}. Kiel anta\u ue, la rigida stango movi\^gas paralele al si, la\u ulonge akso $x$\,.}
\pprn{In the preceding section we studied the motion of a rod which remained rigid under an arbitrary acceleration. We now specialize, choosing $x(t)$ such that the clock $a$ have constant proper acceleration
\cite[page~22]{LL}, \cite[page~73]{Moller}, \cite[page~49]{Rindler}, \cite{GiannoniGron} and \cite{Nikolic}. As before, the rigid rod moves parallel to itself, along the $x$-axis.}

\ppl{La propra akcelo $A$ de partiklo estas {\mbox difinata} kiel $A:=\dd^{\,2}x_v/\dd{t_v}^2$\,, per \^ciu inercia {\mbox sistemo} de referenco $S_v$ momente ripozanta de la {\mbox partiklo}. Pro tiu akcelo la partiklo havas diferencialan mov-ekvacion $\dd u/\dd t=A$ en kiu $u:=v/\sqrt{1-v^2/c^2}$, estante $v:=\dd x/\dd t$ la {\mbox partikla} rapido mezurata per la inercia sistemo de referenco $S_0$, kies koordinatoj estas $x$ kaj $t$. La \^generala solvo por $A=\rm konst$ estas}
\ppr{The proper acceleration $A$ of a particle is defined as $A:=\dd^{\,2}x_v/\dd{t_v}^2$\,, for every inertial reference system $S_v$ in which the particle is momentarily at rest. Under this acceleration the particle has differential equation of motion $\dd u/\dd t=A$ where $u:=v/\sqrt{1-v^2/c^2}$, being $v:=\dd x/\dd t$ the particle speed measured in the inertial reference system $S_0$, whose coordinates are $x$ and $t$. The general solution for $A=\rm const$ is}

\bea                                                                               \label{xacel}
x(t) &=& x_0+\frac{c^2}A\left(\sqrt{1+[A(t-t_0)+u_0]^2/c^2}-\gamma_0\right)\,,
\\                                                                                              
\gamma_0 &:=& \frac{1}{\sqrt{1-v_0^2/c^2}}\,, 
\hspace{4ex} u_0:=v_0\gamma_0\,,
\eea

\ppln{en kiuj $t_0$ markas la akcelan komencon, $x_0:=x(t_0)$ estas la komenca loko, kaj $v_0:=v(t_0)$ estas la komenca rapido. La rapido estas}
\pprn{where $t_0$ marks the beginning of the acceleration, $x_0:=x(t_0)$ is the initial position, and $v_0:=v(t_0)$ is the initial speed. The speed is}

\bea                                                                               \label{vacel}
v(t) = \frac{A(t-t_0)+u_0}{\sqrt{1+[A(t-t_0)+u_0]^2/c^2}}\,,
\eea

\ppln{kaj la propratempo inter la akcela komenco en $t_0$ kaj la momento $t$ estas ricevata per la integro de $\dd\tau=\dd t/\gamma(t) = \dd t\sqrt{1-v(t)^2/c^2}$,}
\pprn{and the propertime between the beginning of the acceleration at  $t_0$ and the moment $t$ is obtained through the integration of $\dd\tau=\dd t/\gamma(t) = \dd t\sqrt{1-v(t)^2/c^2}$,}

\bea                                                                             \label{tauacel}
\tau(t) = \frac{c}{A}\sinh^{-1}\left(\frac{A(t-t_0)+u_0}{c}\right)
-\frac{c}{A}\sinh^{-1}\left(\frac{u_0}{c}\right).
\eea 

\ppl{La horlo\^go $a$ de la anta\u ua sekcio havas $x_0=0$, $v_0=0$, kaj $t_0=0$. La ekvacioj~(\ref{xacel}) -- (\ref{tauacel}) por tiu $a$ simpli\^gas al}
\ppr{The clock $a$ of the preceding section has $x_0=0$, $v_0=0$, and $t_0=0$. The equations~(\ref{xacel}) -- (\ref{tauacel}) for that $a$ simplify to}

\bea                                                                              \label{xacel2}
x &=& \frac{c^2}{A}\left(\sqrt{1+A^2t^2/c^2}-1\right)
= \frac{2c^2}{A}\sinh^2\left(\frac{A\tau}{2c}\right) ,
\\ \label{vacel2}                                                                               
v &=& \frac{At}{\sqrt{1+A^2t^2/c^2}}
= c\,\tanh\left(\frac{A\tau}{c}\right) ,
\\
\gamma &=& \sqrt{1+A^2t^2/c^2}                                                                  
= \cosh\left(\frac{A\tau}{c}\right),
\ \ \ \ \
u = At=c\,\sinh\left(\frac{A\tau}{c}\right),
\\ \label{tauacel2}                                                                             
\tau &=& \frac{c}{A}\sinh^{-1}\left(\frac{At}{c}\right) .
\eea

\ppl{Ni determinas nun la movadon de la horlo\^go $b$, anka\u u fiksita en la stango, en la komenca loko $x=L$. Uzante la anta\u uajn ekvaciojn en la rigidecaj kondi\^coj (\ref{xb}) ni ricevas}
\ppr{We now determine the motion of the clock $b$, also fixed on the rod, at the initial position $x=L$. Using the preceding equations under the rigidity conditions (\ref{xb}) we obtain}

\bea                                                                             \label{xbconst}
x_b &=& L+\frac{c^2}{B}\left(\sqrt{1+B^2t_b^2/c^2}-1\right)
= L+\frac{2c^2}{B}\sinh^2\left(\frac{B\tau_b}{2c}\right),
\\ \label{tbconst}                                                                              
t_b &=& \frac{A}{B}\,t\,, \hspace{4ex} B:=\frac{A}{1+AL/c^2}\,.
\eea

\ppln{Komparante~(\ref{xbconst}) kun~(\ref{xacel2}) ni vidas ke la horlo\^go $b$ ekiras de la loko $L$ kiam $t_b=0$ kaj havas propran akcelon $B$ anka\u u konstanta. Ni rimarkas ke, se la horlo\^go $b$ estus en la komenca loko $-c^2/A$, tial \^gia propra akcelo estus nefinia kaj \^gia rapido estus $c$\,; tio signifas, ke ni ne povas senfine poste plivastigi la stangon. Kaj~(\ref{tbconst}), skribata kiel $Bt_b=At$, permesas tre klaran Newtonan interpreton: se la horlo\^go $b$ ricevas konstantan akcelon $B$ dum la intertempo $t_b$, \^gi havos saman rapidon kiel la horlo\^go $a$, se \^ci tiu ricevas konstantan akcelon $A$ dum la intertempo $t$.}
\pprn{Comparing~(\ref{xbconst}) with~(\ref{xacel2}) we see that the clock $b$ starts from the position $L$ when $t_b=0$ and has proper acceleration  $B$ also constant. We remark that, if the clock $b$ were at the initial position $-c^2/A$, its proper acceleration would be infinite and its speed would be $c$; this means that we can not enlarge the rod indefinitely backwards. And~(\ref{tbconst}), written as $Bt_b=At$, permits a very clear Newtonian interpretation: if the clock $b$ receives constant acceleration $B$ during the time interval $t_b$, it will get the same speed as the clock $a$, if this receives constant acceleration $A$ during the time interval $t$.}

\ppl{Evidente, \^ciu rezulto havigata en sekcio~\ref{rig} estas anka\u u valida en \^ci tiu speciala okazo. Sed ni nun havas unu novan rezulton. En iu ajn momento dum la rigida stango akceli\^gas, la registroj de la horlo\^goj $a$ kaj $b$ vidataj per $S_0$ estas donataj per~(\ref{tauacel2}), kun $B$ anstata\u u $A$ por la kalkulo de $\tau_b$.}
\ppr{Evidently, every result obtained in section~\ref{rig} is also valid in this special case. But we now have one new result. At any moment when the rigid rod is accelerating, the readings of the clocks $a$ and $b$ seen in $S_0$ are given by~(\ref{tauacel2}), with $B$ in place of $A$ for the calculation of $\tau_b$.}

\ppsection[-0.6ex]{Konkludoj}{Conclusions}

\ppln{Ni vidis ke la relativeca tempo estas malsama ol la Newtona: la fluo de la relativeca (propratempo) dependas de la rapido de la korpo (horlo\^go) kiu mezuras \^gin, kontra\u ue la Newtona ne dependas. La\u u la relativeco, ju pli granda la horlo\^ga rapido, des malpli rapida la anta\u ueniro de \^gia registro.}
\pprn{We saw that the relativistic time is different from the Newtonian: the flow of the relativistic (propertime) depends on the speed of the body (a clock) that measures it, while the Newtonian does not depend. According to relativity, the larger the speed of the clock, the less fast the increasing of its register.}

\ppl{Do, ni vidis en subsekcio~\ref{casogeral} ke, kontra\u ue la Newtona anta\u uvido, Dopplera efiko de alru\^go estas ebla e\^c se la interspaco fonto-observanto plieti\^gas. Ni anka\u u vidis, en sekcio~\ref{gemeos}, ke la detala analizo de Darwin montris ke la \^general-relativeco ne estas necesa por solvi la \^gemel-paradokson. Estis same grava noti, en sekcio~\ref{circulo}, ke horlo\^goj fiksitaj en la bordo de rotacia disko ne povas esti \^ciue sinkronaj la\u u Einstein.}
\ppr{So, we saw in subsection~\ref{casogeral} that, oppositely to the Newtonian prediction, a Doppler effect of redshift is possible even if the distance source-observer is decreasing. We also saw, in section~\ref{gemeos}, that the detailed analysis of Darwin showed that the general relativity is not necessary to solve the twin paradox. It was also important to remark, in section~\ref{circulo}, that clocks fixed on the rim of a rotating disk can not be all Einstein synchronous.} 

\ppl{En sekcio~\ref{rig}, ni vidis ke ne estas triviala, la konstruo de movi\^ganta inercia sistemo de referenco ekde iu alia $S$ senmova. Fakte, se ni same akcelas \^ciujn punktojn de $S$ (kiel en sekcio~\ref{Srekto}), ni perdas \^gian rigidecon, kaj ne havigas la spacan malplivasti\^gon nek la malsinkronon ordinarajn de la special-relativeco. Aliaflanke, se ni akcelas $S$ trudante rigidecon, ni havigas la Lorentzan malplivasti\^gon, sed ankora\u u ne ekhavis la volatan malsinkronon. Por fine havigi inercia sistemo de referenco, resinkronigo de movi\^gantaj horlo\^goj estas bezona.} 
\ppr{In section~\ref{rig}, we saw that it is not trivial, the construction of a moving inertial reference system starting from some other  $S$ at rest. In fact, if we accelerate equally all points of $S$ (as in section~\ref{Srekto}), we loose its rigidity, and do not obtain the ordinary spatial contraction and dis-synchronism of special relativity. On the other side, if we accelerate $S$ imposing rigidity, we obtain the Lorentzian contraction, but still do not obtain the desired dis-synchronism. To finally obtain an inertial reference system, a re-synchronization of the moving clocks is necessary.}

\ppl{Fine, en sekcio~\ref{acelconst} ni vidis ke se iu punkto de la rigida stango havas konstantan propran akcelon, do \^ciuj aliaj punktoj anka\u u havos \^gin, sed kun malsamaj valoroj; tiuj ie anta\u uaj en la movado estas la malpli akcelataj, kaj la stango ne povas tro ie poste  plivasti\^gi.} 
\ppr{Finally, in section~\ref{acelconst} we saw that if one point of the rigid rod has constant proper acceleration, then all oter points also have it, but with different values; those that are ahead in the motion are less accelerated, and the rod can not extend backwards too much.}  

\ppl{En inercia sistemo de referenco de la special-relativeco, la valoro de la tempa koordinato estas trudata per horlo\^goj sinkronaj la\u u Einstein. Tamen, ni povas uzi anka\u u neinerciajn sistemojn de referenco, en kiuj la tempa koordinato estas trudata per mezuriloj kies taktoj estas malsamaj ol tiu de la normohorlo\^goj de la special-relativeco. Tio estas temo por nia estonta artikolo, {\em La relativeca tempo~-~II}~\cite{Lrt2}.}
\ppr{In an inertial reference system of the special relativity, the value of the time coordinate is imposed by clocks Einstein synchronized. However, we can use also non-inertial reference systems, in which the time coordinate is imposed by measuring apparatuses whose paces are  different from that of the standard clocks of special relativity. This is the theme of our future article, {\em The relativistic time~-~II}~\cite{Lrt2}.}


\ppdu{\vspace{2em}
\ppl{\section*{~}\vspace{-1em}} \nopagebreak
\ppR{\section*{References}\vspace{-1em}} \ppp \nopagebreak
\vspace{-1.9em}}
\selectlanguage{esperanto}

\ppdu{\end{Parallel}}

\end{document}